\documentclass[11pt]{llncs}

\usepackage{times}
\usepackage{subfigure}
\usepackage{epsfig}
\usepackage{amsmath}
\usepackage{amssymb}

\def\Proof{\par\noindent{\bf Proof:}\indent}
\def\QED{\hfill$\Box$\par\vskip1em}
\def\PROG#1{${\cal #1}$}
\def\MPROG#1{{\cal #1}}

\begin{document}

\title{Ideal Stabilization}

\author{Mikhail Nesterenko\inst{1}\thanks{This
research is supported in part by NSF Career award CNS-0347485.}
\and S\'ebastien Tixeuil\inst{2}\thanks{This author is supported in part by the ANR SHAMAN project.}}
\institute{Kent State University, USA
\and
Universit\'{e} Pierre \& Marie Curie - Paris 6, France}
\maketitle
\begin{abstract}
We define and explore the concept of ideal stabilization.  The program
is ideally stabilizing if its every state is legitimate. Ideal
stabilization allows the specification designer to prescribe with
arbitrary degree of precision not only the fault-free program behavior
but also its recovery operation. Specifications may or may not mention
all possible states. We identify approaches to designing ideal
stabilization to both kinds of specifications. For the first kind, we
state the necessary condition for an ideally stabilizing solution. On
the basis of this condition we prove that there is no ideally
stabilizing solution to the leader election problem. We illustrate the
utility of the concept by providing examples of well-known programs
and proving them ideally stabilizing.  Specifically, we prove ideal
stabilization of the conflict manager, the alternator, the propagation
of information with feedback and the alternating bit protocol.
\end{abstract}

\section{Introduction}\label{SecIntro}

A program is \emph{self-stabilizing}~\cite{D74j,D00b,T09bc} (or just
\emph{stabilizing}) if, regardless of the initial state, it eventually
satisfies its specification. This elegant property enables the program
to recover from transient faults or lack of initialization. During
this stabilization period the program behavior is unpredictable. It is
tempting to attempt to engineer the specification such that the
program behavior during fault-recovery is controlled. For example, the
program starts behaving correctly in no more than ten steps, or
critical messages are never lost. However, one of the features of
classic stabilization is that the program does not have to satisfy the
specification for an arbitrary amount of time. That is, the program is
free to ignore the recovery constrains built into the
specification. 

Another disadvantage of stabilizing programs is their poor
compositional properties. Stabilization programs are usually composed
by layers: an lower level components are not influenced by the higher
level components and, after the lower component starts behaving
correctly, they higher level, due to stabilization will eventually
behave correctly as well. However, if there is non-trivial two-way
interaction between components, the stabilization or correct operation
of the composed system is not guaranteed.  These shortcomings diminish
the attractiveness of stabilization as a viable fault-tolerance
technique.

In this paper we study the class of programs that always satisfy their
specification. We call such programs \emph{ideally
  stabilizing}. Related concepts are periodically considered by
fault-tolerance researchers. However, these approaches are often
regarded as theoretical curiosity with a few isolated examples of no
practical importance. Our thesis is that the opposite is true. Ideal
stabilization retains most of the advantages of classical
stabilization while allowing the engineers and program designers to
control the program behavior during fault recovery. Moreover, since
the ideally stabilizing programs always satisfy their specification,
their composition is similar to conventional program
composition. Thus, the vast array of established program composition
techniques can be applied to ideally stabilizing programs. We are thus
hopeful that our approach to stabilization makes the general concept
of self-stabilization more attractive to fault-tolerance
practitioners.

\paragraph{Related work.} Ideal stabilization is close to
\emph{snap-stabilization}~\cite{BDPV99c,DDNT08c} and should be
considered complementary. However, snap-stabilization is defined in
terms of immediate correct satisfaction of \emph{external}
invocations. Such definition may lead to specifications with
sequence-based safety and liveness properties. Proving stabilization
to such specifications often results in operational proofs that are
difficult to verify. Ideal stabilization, on the other hand, does not
restrict specifications in any way.

Adding safety properties to self-stabilizing protocols has been a very
active recent path of research, however most approaches make some
restrictions on the \emph{nature} or the \emph{extent} of the faults
in order to guarantee a particular kind of safety. Safe
stabilization~\cite{GB03c,BGR06c} refers to the fact that \emph{few}
faults hitting the network should not compromise a particular safety
predicate, while preserving global self-stabilization of the
system. Also, the line of work about fault-containing
stabilization~\cite{GGHP96c} may handle only a \emph{single} transient
failure to ensure actual containment and recovery, and
super-stabilizing protocols~\cite{DH97j} can withstand one topology
change at a time. When faults are simple enough to be detected by
checksums~\cite{HP00j}, it is also possible to maintain elaborate
safety predicates. Other approaches included using \emph{external}
entities that enforce safety~\cite{DS03j} or make use of
non-corruptible memory~\cite{LS92c}. By contrast, ideal stabilization
does not restrict the nature or the extent of the transient faults
that can hit the system in \emph{any} way, and does not make use of
any external entity.

\paragraph{Our contribution.}

In this paper we study two approaches to ideal stabilization. The
approaches depend on the specification type. Specification itself is
ideal if it allows (\emph{i.e.} mentions) all possible
states. Specification is not ideal otherwise. Ideal stabilization to
specifications that are not ideal hinges on the approach we call
\emph{state displacement}. The specification implementer provides such
mapping from program states to specification states that none of the
possible program states map to disallowed specification states.  We
identify the necessary condition for such specifications to allow
ideal stabilization and explain how two well-known programs: conflict
manager and the alternator use state displacement to achieve ideal
stabilization.

The second approach relies on stating the specification such that all
the possible states are allowed. This allows the engineer to specify
precisely what behavior, including failure recovery behavior, is
expected of the program. Ideally stabilizing program, by definition, has
to follow this specification exactly. We state a proposition that
states that such programs are rather common. As an example, we consider
the problem specifications for two well-known stabilizing programs:
propagation of information with feedback and alternating-bit protocol
and provide assertional proofs that the programs ideally stabilize to
these specifications.

\section{Model}\label{SecModel}

This section introduces the notation and terms we use in the
rest of the paper. To the person familiar with the literature on
self-stabilization, our notation may look fairly
conventional. However, we encourage even the specialists to read
this section as the understanding of the results in the
further sections hinges on the notions defined in this one.

\ \\ \textbf{Program.} A \emph{program} consists of a set of $N$
processes. Each process contains a set of variables.  Every variable
ranges over a fixed domain of values. Variable $v$ of process $p$ is
denoted $v.p$. A process \emph{state} is an assignment of a value from
its domain to each variable of the process. A program state, in turn,
is an assignment of a value to every variable of each process. The
Cartesian product of the values of all program variables is program
\emph{state universe}. That is, the state universe defines all states
that the program can assume.

Each process also contains a set of actions. An action has the form
$\langle name\rangle : \langle guard \rangle \longrightarrow \langle
command \rangle$. A \emph{guard} is a predicate over the variables of
the process. A \emph{command} is a sequence of assignment and
branching statements.

%
%
An action whose guard is \textbf{true} at some program state is
\emph{enabled} in that state. The execution of an enabled action
changes the values of program variables and thus transitions the
program from one state to another. A program \emph{computation} is a
maximal sequence of such transitions. By maximality we mean that the
computation is either infinite or it ends in a state where none of the
actions are enabled. Note that we do not assume any fairness of action
execution for infinite computations.

\ \\ \textbf{Communication model, extended state.} Processes that
share variables are \emph{neighbors}.  The communication model
determines the type and access method of the variables shared by
neighbor processes. For example, in \emph{shared-memory} communication
model, the process may mention the variables of the neighbor processes
in its actions. That is, the process may read the state of its
neighbor processes. The \emph{extended} state of the process is the
state of its local variables and the variables that the process can
read.

\ \\ \textbf{Problem specifications and program sequence mapping.}
\emph{Problem specification} prescribes the program behavior. This is
done by defining the program inputs and outputs through \emph{external
  variables}. External variables are thus either \emph{input} or
\emph{output} variables. Input variables are modified by the
environment while the program may only read them. The output variables
are updated by the program, they are used to display the results of
the program computation.

The problem specification is the set of sequences of states of
external variables.  A program implements the specification.  Part of
the implementation is the mapping from the program states to the
specification states. This mapping does not have to be
one-to-one. However, we only consider \emph{unambiguous} programs
where each program state maps to only one specification state. 

%
%
We make another important assumption about program sequence mappings.
The mappings have to be \emph{merge-symmetric}. Specifically, let
there be a set of program states $pr_1$ through $pr_k$ such that they
map to specification states $s_1$ through $s_k$. Then, if there is a
specification state $s_m$ such that the extended state of each process
$p$ in $s_m$ is the same as in one of the states $s_1$ through $s_k$,
then there exists a program state $pr_m$ that maps to $s_m$. In other
words, any specification state formed by extended process
state-preserving combination of other specification states, has a
program state that maps to this new specification state.  Most known
mappings are merge-symmetric.

\emph{Identical} mapping maps every program state to the specification
state.  That is, the program operates external variables only.
Another simple program mapping is \emph{projection}. Each process
maintains output variables and internal variables for computations and
record keeping. The projection of program states onto specification
states removes the internal variables.  However, the mapping may not
be as straightforward as identical mapping or projection. For example,
the specification requires the output variable of a process be boolean
while the program maintains an integer variable. The mapping is such
that the even values of the integer variable are mapped to
\textbf{true} while odd values to \textbf{false}.

Once the mapping between program and specification states is
established, the program computations are mapped to specification
sequences as follows. Each program state is mapped to the
corresponding specification state. Then, \emph{stuttering}, the
consequent identical specification states, is eliminated.

The program does not have to implement all specification
sequences. However, the program cannot select implementation sequences
so that it ignores environment input. Hence the following notion of
input completeness.  Given a set of sequences \PROG{A}, a subset
$\MPROG{B} \subset\MPROG{A}$ is \emph{input-complete} if, for every
sequence $\alpha \in \MPROG{A}$, there exists a sequence $\beta \in
\MPROG{B}$ such that: every step $s_1$ of $\alpha$ is also in $\beta$
and for every pair of steps $s_1$ and $s_2$ their order in $\alpha$
and $\beta$ is the same. Informally, the sequences in an
input-complete subset \PROG{B} preserve the results and the order of
the input steps in \PROG{A}.

The state universe of the specification is the Cartesian product of
the values of all the external variables. The state that is present in
one of the specification sequences is \emph{allowed} by the
specification. The state is \emph{disallowed} otherwise. The
specification is \emph{ideal} if it allows every state in its state
universe; otherwise the specification is not ideal.

We only consider specifications that are suffix-closed. That is, every
suffix of a specification sequence is also a sequence in this
specification. Suffix closure enables us to discuss the correctness of
the program on the basis of its current state rather than potentially
arbitrary long program history. This facilitates assertional reasoning
about program correctness.

\ \\ \textbf{Predicates, invariants, stabilization.} A state
\emph{predicate} is a boolean expression over program variables. A
program state \emph{conforms} to predicate $R$, if $R$ evaluates to
\textbf{true} in this state; otherwise, the state \emph{violates} $R$.
By this definition every state in the program state universe conforms
to predicate \textbf{true} and none conforms to \textbf{false}.

The predicate defines a set of program states that conform to it. In
the sequel we use the predicate and the set of states it defines
interchangeably. Predicate $R$ is \emph{closed} in a certain program
\PROG{P}, if every state of every computation of \PROG{P} conforms to
$R$ provided that the computation starts in a state conforming to $R$.

A closed predicate $I$ is an \emph{invariant} of the program \PROG{P}
with respect to specification \PROG{S} if $I$ has the following
property: every computation of \PROG{P} that starts in a state
conforming to $I$, maps to a sequence that belongs to the
specification.  A program state is \emph{legitimate} if it conforms to
the invariant and \emph{illegitimate} otherwise. 

Program \PROG{P} \emph{satisfies} (or \emph{solves}) specification
\PROG{S}, if there exists an invariant $I$ of \PROG{P} with respect to
\PROG{S} such that the mappings of the program computations that start
from $I$ form an input-complete subset of \PROG{S}. That is, the
program does not have to implement all the specification sequences,
but it does need to accommodate all possible inputs. Specifically, it
needs to implement an input-complete subset of these sequences.

A program \PROG{P} is \emph{stabilizing} to specification \PROG{S} if
every computation that starts in an arbitrary state of the program
sate universe contains a state conforming to the invariant with
respect to \PROG{S}. Therefore, any computation of a stabilizing
program contains a suffix that implements a specification sequence.

\begin{definition}
A program is \emph{ideally} stabilizing if every state in its state
universe is legitimate.
\end{definition}

That is, \textbf{true} is an invariant of an ideally stabilizing
program.

\section{State Displacement}\label{SecNonIdealSpec}

\subsection{Necessary Condition for Specification}

A non-ideal specification disallows certain states in its state
universe. Yet, every state in the program state universe is
legitimate. Thus, for a program to ideally stabilize to such
specification, the state mapping should be such that the disallowed
states are displaced. That is, none of the states in the program state
universe maps to the disallowed states. However, state displacements
may not be possible for an arbitrary specification. In this section we
provide a theorem that establishes a necessary condition for a
specification to be solvable by an ideally-stabilizing program.

\begin{theorem}\label{trmNoIdeal}
An ideal stabilization is possible to non-ideal specification only if
the specification contains an input-complete subset of sequences such
that in every disallowed specification state there is at least one
process whose extended state does not occur in any of the
specification states of this subset.
\end{theorem}

\Proof Assume the opposite. There is, a non-ideal specification
\PROG{S} that disallows state $d$ and for every input-complete subset
\PROG{C} of \PROG{S} and for every process $p_i$, where $i=1,N$, there
is a specification state $c_i$ in one of the sequences of \PROG{C}
such that the extended state of $p_i$ is the same in $c_i$ and in
$d$. However, there is a program \PROG{P} that ideally stabilizes to
\PROG{S}.

Since \PROG{P} solves \PROG{S}, \PROG{P} implements an input
complete-subset of \PROG{S}.  Assume, without loss of generality, that
\PROG{P} implements \PROG{C}. That is, for every sequence of \PROG{C}
there is a computation of \PROG{P} that maps to this sequence. This
means that for each specification state of \PROG{C}, there is a
program state of \PROG{P} that maps to it.  This includes the states
$c_i$. For each $i$, let $pr_i$ be the program state of \PROG{P} that
maps to $c_i$.

Recall that the extended state of each process in $d$ is the same as
in one of the states $c_i$. If this is the case then, according to
merge-symmetry of program mapping, there exists are program state
$pr_d$ that maps to $d$. Let us consider the computation of \PROG{P}
that starts in $pr_d$. This computation contains a state that maps to
a disallowed state, $pr_d$ itself. That is, this computation maps to a
sequence outside the specification \PROG{S}. This means that $d$ does
not conform to an invariant of \PROG{P} with respect to \PROG{S}. That
is, $pr_d$ is an illegitimate state. However, if \PROG{P} has an
illegitimate state then, contrary to our initial assumption, \PROG{P}
does not ideally stabilize to \PROG{S}. Hence the theorem.  \QED

\subsection{Examples}
To illustrate the concept of ideal stabilization to non-ideal
specifications and the ramifications of Theorem~\ref{trmNoIdeal}, we
provide several examples.

\ \\ \textbf{Conflict manager.} The specification we consider is a
simplified (unfair) variant of the dining philosophers
problem~\cite{CM84,D68} that we call \PROG{UDP}. The program is
adapted from the deterministic conflict manager presented by
Gradinariu and Tixeuil~\cite{GT07cb}. The processes are arranged in a
chain. Every process has a unique identifier. The specification
defines one external output boolean variable $in$ per process. If the
value of $in$ is \textbf{true}, the process may execute the exclusive
critical section of code. The specification defines infinite sequences
where $in$ variables alternate between \textbf{true} and
\textbf{false}. The sequences are not necessarily fair as a certain
process may never be given a chance to execute the crucial
section. That is, an input-complete subset of the specification
contains any subset of such sequences.

The specification prohibits concurrent critical section access by
neighbor processes. That is, the specification disallows states where
$in$ variables of two neighbors are \textbf{true} in the same
state. Assuming shared memory communication, the extended state of the
process contains the state of its neighbors. Hence, none of the
allowed specification states contain an extended process state where
both the process and one of its neighbors are inside the critical
section. The specification thus satisfies the conditions of
Theorem~\ref{trmNoIdeal}.

The conflict manager program \PROG{CM} is as follows. Each
process has a single boolean variable \emph{access} and a single
action \emph{flip} that is always enabled. The action toggles the
value of \emph{access}.

\[\emph{flip}: \textbf{true} \longrightarrow access := \neg access \]

The program mapping is this. For each process $p$ the variable $p.in$
is \textbf{true} if $p.access$ is \textbf{true} and $p$ has the
highest identifier among its neighbors with $access$ set to
\textbf{true}.

Let us discuss why this mapping is merge-symmetric.  Any extended
process state-preserving combination of specification states produces
a specification state where neighbors are not accessing the critical
section. Then, by appropriately setting $access$ variables, we can
generate the program state that maps to this specification state.

Let us give an illustration for this reasoning. Assume we have a chain
of four processes with identifiers $\langle 2,1,3,4\rangle$. The
extended state of each process in this case is its own state plus the
state of its left and right neighbors.  Consider two specification
states  
\[s_1 \equiv
\langle\textbf{true}, \textbf{false}, \textbf{false}, \textbf{false}\rangle\]
and 
\[s_2 \equiv
\langle\textbf{false}, \textbf{false}, \textbf{false}, \textbf{true}\rangle\]
Some of the program states that map to $s_1$ and $s_2$ are
respectively $pr_1 \equiv
\langle\textbf{true},\textbf{true},\textbf{false},\textbf{false}\rangle$
and $pr_2 \equiv
\langle\textbf{false},\textbf{false},\textbf{false},\textbf{true}\rangle$.

Specification state $s_3
\equiv\langle\textbf{true},\textbf{false},\textbf{false},\textbf{true}\rangle$
is formed by merging states $s_1$ and $s_2$. Note that the extended
states of each process in $s_3$ are the same in either $s_1$ or
$s_2$. For example, the extended state of process $p_2$ is $\langle
\textbf{true}, \textbf{false}, \textbf{false}\rangle$, which is the
same in $s_1$ and $s_3$. Note that there are a number of program
states that map to $s_3$. For example, $pr_3 \equiv
\langle\textbf{true},\textbf{false},\textbf{false},\textbf{true}\rangle$.

\begin{theorem}\label{trmCM}
Program \PROG{CM} ideally stabilizes to the unfair dining philosophers
specification \PROG{UDP}.
\end{theorem}

\Proof To prove ideal stabilization we need to show that a computation
of \PROG{CM} from an arbitrary program state satisfies
\PROG{UDP}. First, we show that every state of \PROG{CM} maps to an
allowed state of \PROG{UDP}. Indeed, among the neighbors whose
$access$ is \textbf{true}, $in$ is set to \textbf{true} only for the
process with the highest identifier. That is, every program state maps
to the state of \PROG{UDP} where neighbors do not access the critical
section concurrently. Hence, no program state maps to a disallowed
state. 

Moreover in every computation of the program at least one process, the
process with the largest identifier in the system, alternates between
setting \emph{access} to \textbf{true} and \textbf{false}. This means that
this process alternates between entering and exiting the CS
indefinitely. Such computations satisfy the specification. That is,
\PROG{CM} ideally stabilizes to \PROG{UDP}.\QED

\ \\ \textbf{Leader election.} We show a simplified leader election
problem \PROG{LE} as an example of the specification to which ideally
stabilizing solutions do not exist. Again, the processes form a
chain. In this case, we only consider $N > 3$.  In the external state,
each process has two boolean variables: an input variable $contend$
and an output variable $leader$. The value of the input variable
$contend$ is set to a particular value and does not change throughout
the specification sequence.  In each specification state $leader$ of
at most one process is $\textbf{true}$. To exclude trivial solutions,
the specification requires that the leader is elected only out of the
processes that contend for leadership. That is, the processes whose
$contend$ variable is \textbf{true}.  Each specification sequence is
finite and ends with a state where the leader is elected.  Note that
the input complete subset of sequences has to contain a sequence for
every combination of the contending processes.

\begin{theorem}
There does not exist a program that ideally stabilizes to the
simplified leader election specification \PROG{LE}.
\end{theorem}

\Proof Let us consider state $s_1$ where the first process is
the only one contending for leadership. This is the process that has
to be elected leader. That is, the following output state has to be in
every input-complete subset.  $s_1 \equiv
\langle\textbf{true},\textbf{false},\cdots,
\textbf{false},\textbf{false}\rangle$ Similarly, let $s_2$ be the
state where the last process is the only one contending: $s_2 \equiv
\langle\textbf{false},\textbf{false},\cdots,\textbf{false},\textbf{true}\rangle$.

We now form the state $s_3$ where both the first and the forth
processes are contending for leadership and both of them are
elected. That is, 
\[s_3 \equiv
\langle\textbf{true},\textbf{false},\cdots,\textbf{false},\textbf{true}\rangle
\]
This state is disallowed . Yet, the extended state of every process is
present in either $s_1$ or $s_2$. Thus, according to
Theorem~\ref{trmNoIdeal}, ideal stabilization is not possible to
\PROG{LE}. \QED

\ \\ \textbf{Linear alternator.} For another example, we demonstrate
how a well-known program called the linear alternator \PROG{LA}
proposed by Gouda and Haddix~\cite{GH99ca} fits into our ideal
stabilization model. The alternator provides a solution to the fair
variant of the dining philosophers problem \PROG{FDP}. The problem
specification is the same as described above except all the sequences
are fair with respect to the process critical section access. The
modified specification still excludes the states where two neighbors
are executing the critical section concurrently and the specification
still satisfies the conditions of Theorem~\ref{trmNoIdeal}. Therefore,
the ideal solution is still possible for this specification.

The implementation of \PROG{LA} is as follows. Similar to \PROG{CM}
manager, each process has a boolean variable $x$. This time though we
assume that the processes in the chain are numbered in the increasing
order from $1$ to $N$. The numbering is for presentation purposes only
as as each process only need to be aware of its right and left
neighbor. The process actions are as follows. In the
actions, parameter $j$ ranges from $2$ to $N-1$.

\[
\begin{array}{lll}
x.p_1=x.p_2 & \longrightarrow & x.p_1 := \neg x.p_1 \\
(x.p_j \neq x.p_{j-1}) \wedge (x.p_j = x.p_{j+1}) 
  &\longrightarrow & x.p_j := \neg x.p_j \\
x.p_{N-1} \neq x.p_{N} &\longrightarrow & x.p_N := \neg x.p_N 
\end{array}
\]

The program to specification states mapping is as follows. For each
process $p_j$, the output variable $in.p_j$ evaluates to \textbf{true}
if process' action is enabled.


\begin{theorem}
Program \PROG{LA} ideally stabilizes to the fair dining philosophers 
specification \PROG{FDP}.
\end{theorem}
\Proof

Gouda and Haddix~\cite{GH99ca} prove that the alternator satisfies the
fairness properties of the dining philosophers specification from an
arbitrary program state. We only show the displacement of disallowed
states. Note that an action of a process $p$ is enabled if $x.p$ is
not equal to the left neighbor's variable and equal to the right
neighbor's variable. This can only hold for one process in the
neighborhood. That is, every program state maps to a specification
state where none of the neighbors are in the critical sections
concurrently. In other words, the program states only map to the
allowed states.  Hence the theorem. \QED

\section{Stabilizing to Ideal Specifications}\label{SecIdealSpec}

\subsection{Forming Ideal Specifications}
Another method of achieving ideal stabilization is by stating the
specification such that all the states in its universe are
legitimate. That is, stating ideal specification. At first sight this
seems difficult to achieve. However, the following proposition
demonstrates that such specifications are rather common.

\begin{proposition}
For every program there is an ideal specification to which this
program ideally stabilizes.
\end{proposition}

We provide an informal argument for the validity of this
proposition. Consider any program and all the computations produced by
this program when it starts from an arbitrary state of its
universe. Now define the specification that contains exactly these
computations. This specification is ideal as all the states of its
state universe are allowed while the program ideally stabilizes to
this specification.

Naturally, this kind of specification may not be very useful as it,
in essence, defines specification to be whatever the program
computes. However, below we describe how a number of stabilizing
program published in the literature can be defined as ideally
stabilizing to ideal specifications.

\subsection{Examples}

\textbf{Propagation of Information with Feedback.} As the first
example we describe a program \PROG{PIF} that ideally stabilizes to
the propagation of information with feedback~\cite{C82,S83}
specification. The presentation of this program as snap-stabilizing is
well-known~\cite{BDPV99c}. The program is presented on rooted
trees. However, to simplify the presentation, we describe the
operation of this program on a chain.

For ease of exposition we preserve the rooted tree terminology.
Similarly to the alternator, we assume that the processes are numbered
from $1$ to $N$ from left to right in the chain. We refer to the
leftmost processor, with identifier $1$, as \emph{root}; the
rightmost, with identifier $N$, as \emph{leaf}; and the processes in
between as \emph{intermediate}. For each process $p$, the processes to
the left of it are \emph{ancestors} and to the right ---
\emph{descendants}.

Each process has a state variable $st$.  In the intermediate
processes, the variable may hold one of the three values: \textbf{i},
\textbf{rq}, \textbf{rp} which stand for \emph{idle},
\emph{requesting} and \emph{replying} respectively. The root can only
be idle or requesting while the leaf can be either idle or replying.

The objective of the program is to send a signal from the root to the
leaf and in return receive an acknowledgment that matches this
signal. Operationally, the program should ensure that after the root
makes a request then every process propagates this request along the
chain from left to right in causally ordered steps transitioning from
idle to requesting. Afterwards, the processes propagate the reply from
right to left in causally ordered steps transitioning from requesting
to replying.

We define the following state predicates. $RP(k)$ are the
specification states where all $k$ processes on the left are
requesting $(k=1,N-1)$ and the rest of the processes are
replying. $RQ(l,m)$ are the specification states where all $l$
processes on the left are requesting $(l=0,N-1)$ and all $m-l$
processes following them are idle $(m = l+1, N)$ while the remaining
processes are replying. Specification \PROG{SPIF} includes the
sequences where the system satisfies one of the predicates and
transitions from one to the other infinitely.

Let us define another pair of predicates. Predicate $RP'(k)$ defines
the states where $k$ processes on the left are requesting $(k=1,N-1)$,
the process $k+1$ is replying and none of the rest are idle. Notice
that $RP(k) \subset RP'(k)$. Predicate $RQ'(l,m)$ are the states where
all $l$ processes on the left are requesting $(l=0,N-1)$, all $m-1$
following them are idle $(m = l+1, N)$ and the state of the other
processes is arbitrary. Similarly, $RQ(l,m) \subset
RQ'(l,m)$. Specification \PROG{IPIF} includes the sequences where the
system always satisfies either $RP'(k)$ or $RQ'(l,m)$, each sequence
has a sequence in \PROG{SPIF} as a suffix and the transition from
$RQ'(l,m)$ is only to $RP(k)$.

We now describe program \PROG{PIF}. It has only external
variables. The mapping between the program and specification variables
is identical. The actions of \PROG{PIF} are shown in
Figure~\ref{figPIF}.

\begin{figure}
\[
\begin{array}{llllllll}
\emph{request}:& & &st.p_1 = \textbf{i} &\wedge& st.p_2=\textbf{i} 
   & \longrightarrow& st.p_1 := \textbf{rq} \\
\emph{clear}:  & & &st.p_1 = \textbf{rq} &\wedge& st.p_2 = \textbf{rp}
    & \longrightarrow& st.p_1 := \textbf{i} \\
\emph{forward}:& st.p_{j-1} = \textbf{rq} &\wedge& st.p_{j}=\textbf{i} &\wedge& st.p_{j+1} = \textbf{i} 
   &\longrightarrow& st.p_j :=\textbf{rq}\\
\emph{back}:   & st.p_{j-1} = \textbf{rq} &\wedge& st.p_{j}=\textbf{rq} &\wedge& st.p_{j+1}=\textbf{rp} 
   &\longrightarrow& st.p_{j} :=\textbf{rp}\\
\emph{stop}:& st.p_{j-1} = \textbf{i} &\wedge& st.p_{j} \neq \textbf{i} & & 
   &\longrightarrow & st.p_j :=\textbf{i} \\
\emph{reflect}:& st.p_{N-1}=\textbf{rq} &\wedge& st.p_N = \textbf{i} &&
   &\longrightarrow & st.p_{N} := \textbf{rp} \\
\emph{reset}:&   st.p_{N-1} = \textbf{i} &\wedge& st.p_N = \textbf{rp} && 
   &\longrightarrow & st.p_{N} := \textbf{i}
\end{array}
\]
\caption{\PROG{PIF} program actions. Actions \emph{request} and
  \emph{clear} belong to the root process; actions \emph{forward},
  \emph{back}, and \emph{stop} -- to intermediate processes, i.e
  parameter $j$ ranges from $2$ to $N-1$; actions \emph{reflect} and
  \emph{reset} -- to the leaf.  }\label{figPIF}
\end{figure}

\begin{theorem}
\PROG{PIF} classically stabilizes to \PROG{SPIF} and ideally
stabilizes to \PROG{IPIF}.
\end{theorem}

\Proof First, we prove that $RQ(l,m) \vee RP(k)$ is an invariant of
\PROG{PIF} with respect to \PROG{SPIF}.  Notice that this disjunction
is closed in \PROG{PIF}. That is, none of the actions of \PROG{PIF}
violate the predicate. Let us show that a program transitions from
$RQ(l,m)$ to $RP(k)$ and back. If the program satisfies $RQ(l,m)$, at
least one action is enabled. Indeed, if $l=0, m<N$, \emph{request} is
enabled in \emph{stop} is enabled in process $p_{m+1}$. If,
$l<m-1,m\leq N$, \emph{request} is enabled in $p_{l+1}$. With the
execution of an action either $l$ or $m$ are incremented. If $l=N-1,
m=N$, \emph{reflect} is enabled that transitions the program into
$RP(N-1)$. Similarly, if the system satisfies $RP(k)$ for $k=2,N-1$,
action \emph{back} is enabled. The execution of this action keeps the
system in $RP(k)$ but decrements $k$. If $k=1$, \emph{clear} is
enabled. Its execution moves the program back in $RQ(l,m)$. That is,
the disjunction of the two predicates is closed and a \PROG{PIF}
computation that starts in a state conforming to one of them,
satisfies \PROG{SPIF}. Hence, $RQ(l,m) \vee RP(k)$ is an invariant of
\PROG{SPIF}.

Observe that the disjunction of $RQ'(l,m) \vee RP'(k)$ contains the
program, as well as specification, state universe. We show that
\PROG{PIF} stabilizes to \PROG{SPIF} from this predicate as required
by \PROG{IPIF}.  An argument similar to the above demonstrates that an
action is enabled if \PROG{PIF} satisfies $RP'(k)$.  This action
either decrements $k$ or moves the system to $RQ'(l,m)$.  Also,
similarly, if \PROG{PIF} satisfies $RQ'(l,m)$, then an action is
enabled that increments either $l$ or $m$. Moreover, if $l=N-1,m-N$,
the execution of \emph{reflect} transitions the system to $RP(k)$.
That is, \PROG{PIF} moves from $RP'(k)$ to $RQ'(l,m)$ to $RP(k)$. This
means that the program stabilizes to \PROG{SPIF} and ideally
stabilizes to \PROG{IPIF}.  \QED

\ \\ \textbf{Alternating bit protocol.} Alternating bit protocol is an
elementary data-link network protocol. There is a number of classic
stabilizing implementations of the protocol. Refer to Howell et
al~\cite{HNM99c} for an extensive list of citations.  There is also a
snap-stabilizing version~\cite{DDNT08c}.

The problem is stated as follows. There are two processes:
\emph{sender} --- $p$, and \emph{receiver} --- $q$. The processes
maintain boolean sequence numbers $ns.p$ and $nr.q$. The processes
exchange messages over communication channels. The channels are
reliable and their capacity is one. That is, if the channel is empty,
the message is reliably sent. If the channel already contains a
message, an attempt to send another message leads to the loss of the
new message.  The processes exchange two types of messages: $data$ and
$ack$. Both carry the sequence numbers.

Specification \PROG{SABP} prescribes infinite sequences of states
where there is exactly one message in the channels. The message
carries the sequence number of the sender. The state transitions are
such that $p$ changes the value of $ns$. This change is followed by
the change of the value in $nr$ that matches the value of $ns$.

Specification of \PROG{IABP} is such that for every state in the
universe there is a sequence that starts in it and every sequence
contains a sequence of \PROG{SABP} as a suffix.  Specification of
\PROG{IABP} is thus ideal.

\begin{figure}[ht]
\begin{minipage}[b]{0.5\linewidth}
\begin{tabbing}
12346789\=1234\=1234\=\kill
\emph{next}: \> $\textbf{receive}\ ack(nm) \longrightarrow$ \\
\>\>           $\textbf{if}\ nm = ns\ \textbf{then} $ \\
\>\>\>             $ns:=\neg ns$ \\
\>\>\>            $\textbf{send}\ data(ns)$ \\
\emph{timeout}: \> $\textbf{timeout}() \longrightarrow \textbf{send}\ data(ns)$\\
\emph{reply}: \> $\textbf{receive}\ data(nm) \longrightarrow$ \\
\>\>         $\textbf{if}\ nm \neq nr\ \textbf{then}$\\ 
\>\>\>        $nr:= nm$ \\
\>\>          $ \textbf{send}\ ack(nm) $
\end{tabbing}
\caption{\PROG{ABP} actions.}
\label{figABPcode}
\end{minipage}
\hspace{0.1cm}
\begin{minipage}[b]{0.5\linewidth}
\centering
\epsfig{figure=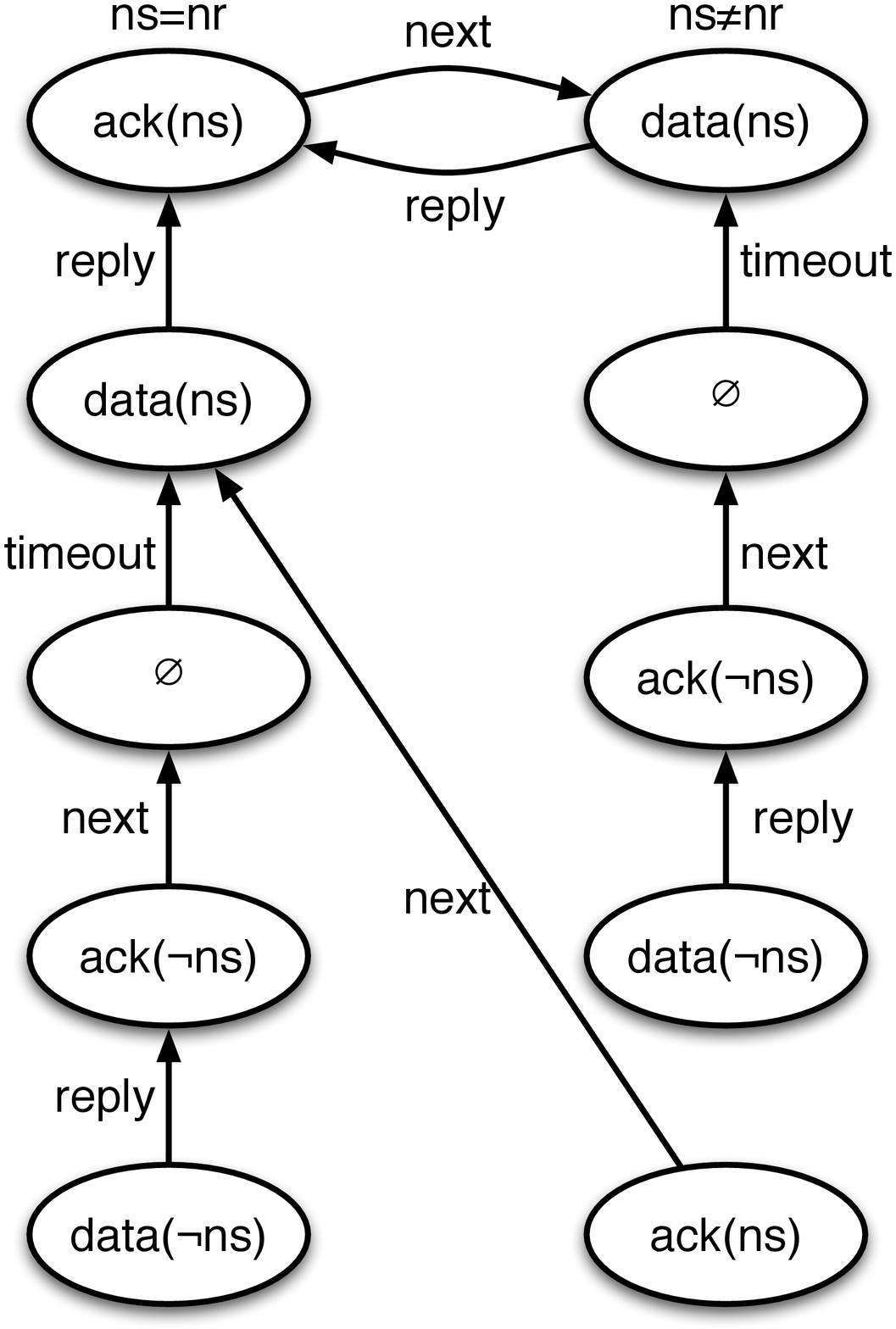,width=4.8cm,clip=}
\caption{\PROG{ABP} state transitions.}
\label{figABPstates}
\end{minipage}
\end{figure}

The program \PROG{ABP} uses only external variables as described by
\PROG{SABP} and \PROG{IABP}. \PROG{ABP} actions are shown in
Figure~\ref{figABPcode}. The sender has two actions: \emph{next} and
\emph{timeout}. Action \emph{next} is enabled if there is a message
from $q$ in the channel. The timeout action is enabled if there are no
messages in either channel.  Upon receiving a message from $q$ with
matching sequence number, $p$ increments the sequence number and sends
the next message. If $p$ times out, it resubmits the same message.
The receiver has a single action. When $q$, receives a message, it
sends an acknowledgment back to $p$. If the message bears a sequence
number different from $rn$, $q$ increments $rn$ signifying the
successful receipt of the message.

\begin{theorem}\PROG{ABP} classically stabilizes to \PROG{SABP} and 
ideally stabilizes to \PROG{IABP}.
\end{theorem}

\Proof We prove the correctness of the theorem by enumerating the
state transitions of \PROG{PIF}.  We classify the state universe of
\PROG{ABP} into two groups: (i) $ns$ is equal to $nr$ and (ii) $ns$ is
not equal to $nr$. The states are further classified according to the
type of messages in the channels. The states and state transitions are
shown in Figure~\ref{figABPstates}. Note that to simplify the diagram
we do not show the states that contain more than a single
message. However, after a single transition, the program moves from
one of those states to a state shown in the figure. The correctness of
the theorem claims can be ascertained by examining the states and
transitions shown in the figure.  \QED

\section{The Impact of Ideal Stabilization Approach}\label{SecEnd}

In this paper we proposed a new way of approaching stabilization. Our
approach eliminates two of the most problematic features of classic
stabilization: unpredictable behavior during stabilization and poor
composability. We hope that this work adds more credence to
stabilization as a viable fault-tolerance technique and generates more
interest in the subject among both theoretical researches and
reliability engineers.

%
%

\bibliographystyle{plain} \bibliography{ideal,../../biblio/biblio}

\end{document}